\documentclass[onecolumn,preprintnumbers,elsart]{revtex4}
\usepackage{amsmath}

\usepackage{mathrsfs}
\usepackage{graphicx}
\usepackage{dcolumn}
\usepackage{bm}
\usepackage[center]{subfigure}

\begin{document}

\title{Discrete solitons and scattering of lattice waves in guiding arrays
with a nonlinear $\mathcal{P}\mathcal{T}$-symmetric defect}
\author {Xiangyu Zhang$^{1}$}
\author{Jinglei Chai$^{1}$}
\author{Jiasheng Huang$^{1}$}
\author{Zhiqiang Chen$^{1}$}
\author{Yongyao Li$^{1}$}
\email{yongyaoli@gmail.com}
\author{Boris A. Malomed$^{2}$}

\affiliation{$^{1}$Department of Applied Physics, South China
Agricultural University, Guangzhou 510642, China \\
$^{2}$ Department of Physical Electronics, School of Electrical
Engineering, Faculty of Engineering, Tel Aviv University, Tel Aviv
69978, Israel.}

\begin{abstract}
Discrete fundamental and dipole solitons are constructed, in an exact
analytical form, in an array of linear waveguides with an embedded $\mathcal{%
PT}$-symmetric dimer, which is composed of two nonlinear waveguides carrying
equal gain and loss. Fundamental solitons in \textit{tightly knit} lattices,
as well as all dipole modes, exist above a finite threshold value of the
total power. However, the threshold vanishes for fundamental solitons in
\textit{loosely knit} lattices. The stability of the discrete solitons is
investigated analytically by means of the Vakhitov-Kolokolov (VK) criterion,
and, in the full form, via the computation of eigenvalues for perturbation
modes. Fundamental and dipole solitons tend to be stable at smaller and
larger values of the total power (norm), respectively. The increase of the
strength of the coupling between the two defect-forming sites leads to
strong expansion of the stability areas. The scattering problem for linear
lattice waves impinging upon the defect is considered too.
\end{abstract}

\maketitle






\section{Introduction}

Arrays of evanescently coupled waveguides made of nonlinear materials is the
fundamental model of discrete nonlinear optics \cite{Lederer,Christodoulides,Garanovich,ZGC}. Guided propagation of light in
such arrays emulates electronic wave functions in fundamental periodic and
disordered potentials of solid state physics. Therefore, discrete arrays of
optical waveguides may be used to implement photonic counterparts of
semiconductor devices used in electronic circuits \cite{Christodoulides,Christodoulides2}. On
the other hand, the flexibility in the creation of virtual (photoinduced)
\cite{Moti1,Moti2} and permanent \cite{Jena1,Jena2,Jena3} arrayed waveguides opens the way to
exploration of phenomena which are difficult to directly observe or control
in solid-state settings, such as the Anderson localization \cite{Anderson1,Anderson2}
and eigenmodes of quasicrystal potentials \cite{quasicryst}. Furthermore,
the use of the material Kerr nonlinearity makes it possible to predict \cite{vortex1,vortex2,vortex3,vortex4,vortex5} and experimentally create \cite{vortex-exper1,vortex-exper2,vortex-exper3} nonlinear modes in
the waveguiding lattices -- notably, various species of discrete vortex
solitons.

The light propagation in arrays with embedded defects has also drawn a great
deal of interest, as the use of defects can enhance the functionally of such
transmission schemes \cite{Peschel,Morandooti,Morales,Molina1,Molina2,Yongyao}. In particular, defects
carrying the balanced gain and loss, which give rise to the optical
realization of the $\mathcal{PT}$ symmetry \cite{PT1,PT2,PT3,PT4,PT5}, were recently
introduced too \cite{Dmitriev1,Dmitriev2,Dmitriev3}. Such systems, although governed by discrete
nonlinear Schr\"{o}dinger equations corresponding to non-Hermitian
Hamiltonians \cite{Zezyulin}, give rise to entirely real propagation
spectra, provided that the common strength of the gain and loss terms does
not exceed a critical level, past which the $\mathcal{PT}$-symmetry of the
propagating eigenmodes is destroyed. The simplest version of discrete $%
\mathcal{PT}$-symmetric nonlinear systems amounts to dimers, i.e., systems
of two linearly-coupled sites, which carry the balanced gain and loss,
combined with the onsite nonlinearity \cite{dimer1,dimer2,dimer3,dimer4}.

In this work, we consider one-dimensional arrays of linear waveguides with
an embedded defect, formed by a pair of Kerr-nonlinear guiding cores with a
tunable strength of the coupling between them. The two defect-forming
waveguides carry mutually balanced linear gain and loss, thus realizing the $%
\mathcal{PT}$ symmetry in this context (a similar configuration built in
continuous medium is called a $\mathcal{PT}$\textit{-symmetric dipole }\cite%
{Mayteevar}). This setting, as well as a similar one, with a pair of
nonlinear waveguides side-coupled to the linear array \cite{side-coupled1,side-coupled2,side-coupled3},
which can be readily implemented in the experiment, offer an advantage of
finding symmetric \cite{embedded1,embedded2} and asymmetric \cite{asymm} solutions for
discrete solitons in an exact analytical form. However, the $\mathcal{PT}$%
-symmetric extension of the dual-core nonlinear defect embedded into a
linear array was not addressed in previous works, although it can also be
realized experimentally, by means of the methods recently developed in
optics \cite{PT1,PT2,PT3,PT4,PT5}. We produce both exact analytical and numerical solutions
for stable fundamental and dipole-mode discrete solitons in this system
(asymmetric solitons cannot be supported by the $\mathcal{PT}$-symmetric
defect, as they cannot maintain the equilibrium between the gain and loss).
In particular, we conclude that the stability and instability regions for
the discrete solitons is accurately identified by the Vakhitov-Kolokolov
(VK) criterion.

The rest of the paper is structured as follows. The system is introduced in
Sec. II. Analytical and numerical results are presented in Sec. III. {The
scattering of lattice waves on the }$\mathcal{PT}$-symmetric defect is
studied by means of direct simulations{\ in Sec. IV} The paper is concluded
by Sec. V.

\section{The Model}
\begin{figure}[tbp]
\centering{\label{fig1a} \includegraphics[scale=0.28]{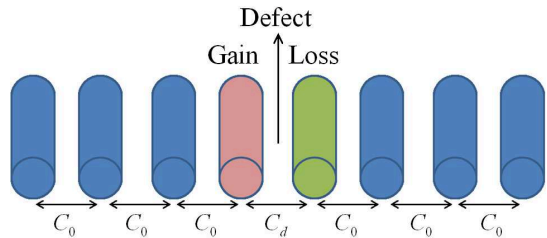}}
\caption{The structure of the system. A pair of nonlinear
cores, carrying mutually balanced gain and loss, are inserted into the host
array of linear waveguides with coupling constant $C_{0}$. The strength of
the coupling between the inserted elements, $C_{d}$, is used as a basic
control parameter.}
\label{Model}
\end{figure}

The system considered in this work is displayed in Fig. \ref{Model}, that
features the one-dimensional linear waveguide array with the embedded pair
of cores, which, in addition to the Kerr nonlinearity, supply equal amounts
of local gain and loss. The nonlinearity can be induced in the inserted
cores by means of appropriate dopants \cite{Kip}, or by making the light
confinement in them much tighter than in the (quasi-)linear waveguides. The
gain and loss may be imposed by means of doping too, with the external pump
focused solely on the core which is selected to carry the gain.

The separation between the embedded cores is considered as a control
parameter of the system. In other words, outside of the defect, the
inter-core coupling coefficient is constant $C_{0}$, while for the two cores
forming the defect it is $C_{d}$. It is implied that $C_{d}/C_{0}>1$ and $%
C_{d}/C_{0}<1$ correspond, respectively, to the distance between the two
defect-building waveguides which is smaller or larger than the separation
between the cores in the linear array.

The propagation of light in the present system is governed by the discrete
nonlinear Schr\"{o}dinger equation, which can be derived by means of
well-known methods (the tight-binding approximation) \cite{Lederer}:
\begin{equation}
i{\frac{du_{n}}{dz}}=-\left( C_{n,n-1}u_{n-1}+C_{n+1,n}u_{n+1}\right)
-\gamma _{n}|u_{n}|^{2}u_{n}+i\kappa _{n}u_{n}.  \label{DNLSE}
\end{equation}%
Here $z$ is the propagation distance, $u_{n}(z)$ is the amplitude of the
electromagnetic field in the $n$-th core, real $C_{n,n\pm 1}$ are the
above-mentioned coupling constants, while real coefficients $\gamma _{n}$
and $\kappa _{n}>0/\kappa _{n}<0$ account for the nonlinearity, and for the
gain/loss constants in the defect-forming cores:
\begin{gather}
C_{n,n-1}=%
\begin{cases}
C_{d}, & n=N/2 \\
C_{0}, & n\neq N/2%
\end{cases}%
,\gamma _{n}=%
\begin{cases}
\gamma , & n=N/2,N/2+1 \\
0, & n\neq N/2,N/2+1%
\end{cases}%
,  \notag \\
\kappa _{n}=%
\begin{cases}
\kappa , & n=N/2 \\
-\kappa , & n=N/2+1 \\
0, & n\neq N/2,N/2+1%
\end{cases}%
,  \label{kappa}
\end{gather}%
where even $N$ is the total number of the waveguides in the system. The
total field power is
\begin{equation}
P=\sum_{n=1}^{N}|u_{n}|^{2}.  \label{P}
\end{equation}

Stationary solutions to Eq. (\ref{DNLSE}) are looked for as
\begin{equation}
u_{n}(z)=U_{n}e^{iKz},  \label{uU}
\end{equation}%
where $U_{n}$ is the distribution of the local amplitudes, and $K$ the
propagation constant. Stability of the localized stationary modes was
investigated numerically by means of computing eigenvalues for small
perturbations, and the results were verified by means of direct simulations
of the perturbed evolution. The perturbed solution was taken as
\begin{equation}
u_{n}=e^{iKz}(U_{n}+w_{n}e^{iGz}+v_{n}^{\ast }e^{-iG^{\ast }z}),  \notag
\end{equation}%
where the asterisk stands for the complex conjugate. The substitution of
this expression into Eq. (\ref{DNLSE}) and linearization leads to the
eigenvalue problem for the perturbation frequency, $G\equiv G_{\mathrm{r}%
}+iG_{\mathrm{i}}$, and the eigenmodes, $\left\{ w_{n},v_{n}\right\} $:
\begin{equation}
\left(
\begin{array}{cc}
C+K+2\gamma _{n}|U_{n}|^{2}+i\kappa _{n} & \gamma _{n}U_{n}^{2} \\
-\gamma _{n}U_{n}^{\ast 2} & -C-K-2\gamma _{n}|U_{n}|^{2}+i\kappa _{n}%
\end{array}%
\right) \left(
\begin{array}{c}
w \\
v%
\end{array}%
\right) =G\left(
\begin{array}{c}
w \\
v%
\end{array}%
\right) .  \label{eigen}
\end{equation}%
Solution $U_{n}$ is stable if all the eigenvalues $G$ are real. In the next
section, we carry out the analytical and numerical study of the stationary
discrete solitons in the present system. By means of rescaling, we set $%
C_{0}=0.5$, $\gamma =1$, and produce numerical results for the system of
size $N=64$, while $P$ [the total power (\ref{P})], $C_{d}$ and $\kappa $
are left as control parameters.

\section{Analytical and numerical results}
\subsection{The analytical consideration}

Following \cite{asymm,Malomed} (and a similar approach for the
continuous system developed in \cite{Mayteevar}), an exact solution for
stationary amplitudes $U_{n}$ with propagation constant $K$ in Eq. ({\ref{uU}%
)} can be sought for as
\begin{equation}
U_{n}=%
\begin{cases}
Ae^{-i\phi /2}\exp \left[ -\lambda \left( {N/2}-n\right) \right] , & \mathrm{%
at}\quad n\leq {N/2,} \\
Ae^{i\phi /2}\exp \left[ -\lambda \left( n-1-{N/2}\right) \right] , &
\mathrm{at}\quad n\geq {N/2}+1,%
\end{cases}
\label{analyt}
\end{equation}%
where $K$ and $\lambda $ are linked by the dispersion relation
\begin{equation}
K=2C_{0}\cosh (\lambda )  \label{DR}
\end{equation}%
(hence, for given $C_{0}$, the propagation constant may only take values $%
K>2C_{0}$), amplitude $A$ may be assumed real, and $\phi $ is, in this form,
an arbitrary phase shift. The substitution of Eqs. (\ref{uU}) and (\ref%
{analyt}) into Eq. (\ref{DNLSE}) at $n=N/2$ and $n=N/2+1$ leads to the final
system of nonlinear equations:
\begin{eqnarray}
&&K=C_{0}e^{-\lambda }+C_{d}e^{+i\phi }+A^{2}-i\kappa ,  \notag \\
&&K=C_{0}e^{-\lambda }+C_{d}e^{-i\phi }+A^{2}+i\kappa ,  \label{K-phi}
\end{eqnarray}%
where, as said above, $\gamma =1$ is fixed. The balance of imaginary terms
in Eq. (\ref{K-phi}) yields two solutions for phase shift $\phi $,
\begin{equation}
\phi _{+}=\arcsin (\kappa /C_{d}),\quad \phi _{-}=\pi -\arcsin (\kappa
/C_{d}),  \label{phi}
\end{equation}%
hence the solution exists under the constraint of $\kappa <C_{d}$. As usual,
this means that $\mathcal{PT}$-symmetric solutions exist provided that the
gain/loss coefficient does not exceed \ a certain critical value. Then, the
balance of real terms in Eq. (\ref{K-phi}) yields an expression for the
squared amplitude:
\begin{equation}
A_{\pm }^{2}=C_{0}e^{\lambda }\mp \sqrt{C_{d}^{2}-\kappa ^{2}},  \label{A^2}
\end{equation}%
where the upper and lower signs correspond, respectively, to $\phi _{+}$ and
$\phi _{-}$ in Eq. (\ref{phi}), and dispersion relation (\ref{DR})\ was used
to simplify this expression. Further, solution (\ref{A^2}) with the lower
sign exists at all values of $\lambda \geq 0$, the limit case of $\lambda =0$
corresponding not to a localized mode, but to a flat one with a constant
amplitude,%
\begin{equation}
A_{-}^{2}\left( \lambda =0\right) =C_{0}+\sqrt{C_{d}^{2}-\kappa ^{2}}.
\label{lambda=0-}
\end{equation}%
As concerns solution (\ref{A^2}) with the lower sign, they exist up to $%
\lambda =0$, also going over into a flat mode with a constant amplitude%
\begin{equation}
A_{+}^{2}\left( \lambda =0\right) =C_{0}-\sqrt{C_{d}^{2}-\kappa ^{2}},
\label{lambda=0+}
\end{equation}%
if the linear lattice is a \textit{tightly knit} one, with the coupling
constant which large enough,%
\begin{equation}
C_{0}>C_{0}^{(0)}\equiv \sqrt{C_{d}^{2}-\kappa ^{2}}  \label{C0}
\end{equation}%
(obviously, the lattice with $C_{0}\geq C_{d}$ is always a tightly knit
one). In fact, both flat states, given by Eqs. (\ref{lambda=0-}) and (\ref%
{lambda=0+}), are unstable because, as shown below, they are limit forms of
broad discrete solitons which are unstable according to the VK criterion.

On the other hand, for a \textit{loosely knit} linear lattice, with $%
C_{0}<C_{0}^{(0)}$, solutions (\ref{A^2}) with the upper sign exist above a
finite value of $\lambda $,%
\begin{equation}
\lambda \geq \lambda _{\min }=(1/2)\ln \left( C_{d}^{2}-\kappa ^{2}\right)
-\ln C_{0}\text{.}  \label{lambda_min}
\end{equation}%
In other words, width $W$ of the respective discrete soliton pinned to the $%
\mathcal{PT}$-symmetric nonlinear defect takes values smaller than the
respective maximum value: $W\equiv 1/\lambda \leq W_{\max }\equiv 1/\lambda
_{\min }$. Note that, unlike amplitudes (\ref{lambda=0-}) and (\ref%
{lambda=0+}) of the flat states, which remain finite at amplitude at $%
\lambda =0$, $A_{+}^{2}(\lambda )$ vanishes at $\lambda =\lambda _{\min }$.

\begin{figure}[tbp]
\centering{\label{fig2a}
\includegraphics[scale=0.2]{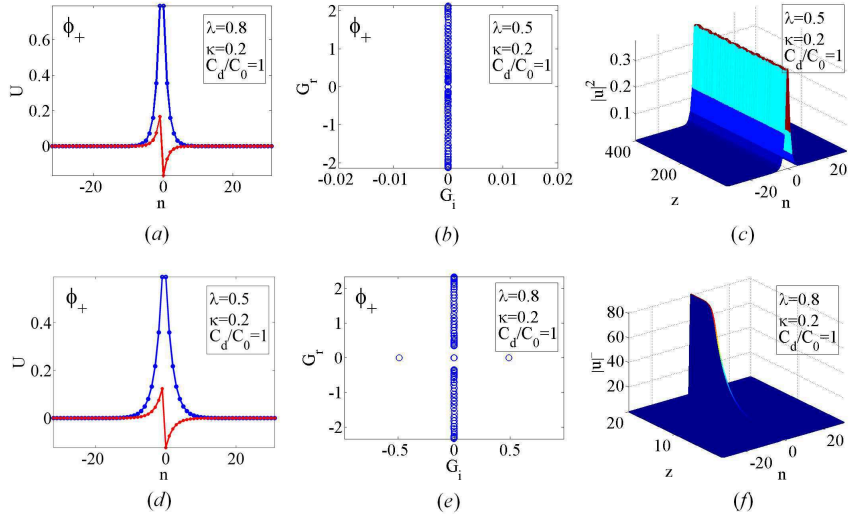}}%
\caption{(a) The blue (even) and red (odd) profiles display
the real and imaginary parts of the stationary lattice field, $U_{n}$ [see
Eq. (\protect\ref{analyt})], of a stable fundamental $\mathcal{PT}$ soliton
for $(C_{d},\protect\lambda ,\protect\kappa )=(0.5,0.5,0.2)$, corresponding
to $\protect\phi _{+}$ and $A_{+}^{2}$ in Eqs. (\protect\ref{phi}) and (%
\protect\ref{A^2}). The total power of this soliton is $P=1.1583$, see Eq. (%
\protect\ref{P+-}). (b,c) Eigenvalues of small perturbations around this
soliton and direct simulations of the perturbed evolution prove that it is
stable. (d) An unstable fundamental $\mathcal{PT}$ soliton solution for $%
(C_{d},\protect\lambda ,\protect\kappa )=(0.8,0.5,0.2)$, with total power $%
P=1.6402$. (e,f) Eigenvalues and direct simulations demonstrate that the
latter soliton is unstable. }
\label{Fundsolution}
\end{figure}

\begin{figure}[tbp]
\centering{\label{fig7a}
\includegraphics[scale=0.2]{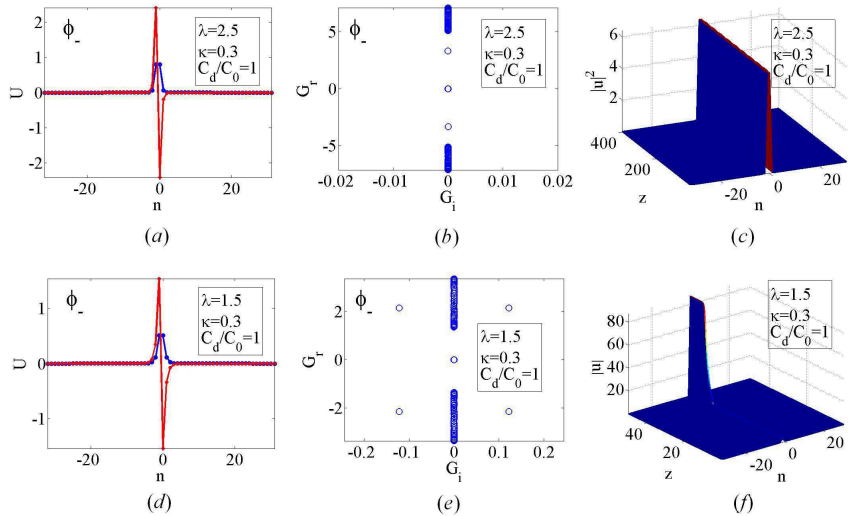}}%
\caption{(a) A stable dipole-mode $\mathcal{PT}$ soliton for $%
(C_{d},\protect\lambda ,\protect\kappa )=(0.5,2.5,0.3)$, corresponding to $%
\protect\phi _{-}$ and $A_{-}^{2}$ in Eqs. (\protect\ref{phi}) and (\protect
\ref{A^2}). The total power of this soliton is $P=13.071$, see Eq. (\protect
\ref{P+-}). (b,c) Eigenvalues and direct simulations demonstrate that this
soliton is stable. (d) An unstable dipole $\mathcal{PT}$ soliton, found for $%
(C_{d},\protect\lambda ,\protect\kappa )=(0.5,1.5,0.3)$, with total power $%
P=5.5584$. (e,f) Eigenvalue and direct simulations demonstrate instability
of the latter soliton.}
\label{Dipolesolution}
\end{figure}

Figures \ref{Fundsolution} and \ref{Dipolesolution} show typical examples of
stable and unstable solutions produced, severally, by Eq. (\ref{analyt})
with $\phi _{+}$ and $\phi _{-}$ (blue and red lines display $\mathrm{Re}%
\left\{ U_{n}\right\} $ and $\mathrm{Im}\left\{ U_{n}\right\} $,
respectively). The profiles of the solutions in Fig. \ref{Fundsolution} show
that the soliton corresponding to $\phi _{+}$ and $A_{+}^{2}$ in Eqs. (\ref%
{phi}) and (\ref{A^2}) may be identified as a fundamental one, with an even
shape of the taller (real) component, while the profiles in Fig. \ref%
{Dipolesolution} demonstrate that the soliton corresponding to $\phi _{-}$
and $A_{-}^{2}$ represents an excited state in the form of a dipole, whose
taller (imaginary) component is odd, with respect to the lattice coordinate.
Unstable solitons undergo a blowup in the real-time propagation, see Figs. %
2(f) and 3(f).

Two solutions (\ref{A^2}) give rise to the following expression for the
total power of the soliton, calculated as per Eq. (\ref{P}):
\begin{equation}
P_{\pm }=2{\frac{C_{0}e^{\lambda }\mp \sqrt{C_{d}^{2}-\kappa ^{2}}}{%
1-e^{-2\lambda }}}.  \label{P+-}
\end{equation}%
In particular, the total power diverges for the dipole modes in the limit of
$\lambda \rightarrow 0$, when the soliton degenerates into the flat state
with constant amplitude (\ref{lambda=0-}), and similarly for the fundamental
mode in the tightly knit lattice, with $C_{0}>C_{0}^{(0)}$, see Eq. (\ref{C0}%
). In these cases, the power attains a finite minimum (threshold), given by
Eq. (\ref{P+-}) at values $\lambda =\lambda _{0}$ determined by a cubic
equation,
\begin{equation}
C_{0}\left( e^{\lambda _{0}}\right) ^{3}-3C_{0}e^{\lambda _{0}}\pm 2\sqrt{%
C_{d}^{2}-\kappa ^{2}}=0.  \label{lambda0}
\end{equation}
On the other hand, for the fundamental mode in the loosely knit lattice,
with $C_{0}<C_{0}^{(0)}$, where the solution family starts from point (\ref%
{lambda_min}), the $P(\lambda )$ dependence commences from $P(\lambda
=\lambda _{\min })=0$.

It is reasonable to expect that a necessary stability condition for the
fundamental and dipole-mode solution families may be given by the
Vakhitov-Kolokolov (VK) criterion \cite{VK1,VK2,VK3}, $dP/dK>0$, or, which is more
technically convenient, $dP/d(e^{\lambda })>0$, in terms of Eqs. (\ref%
{analyt}) and (\ref{DR}). The application of the VK criterion in this form
to expression (\ref{P+-}) leads to the following stability condition:
\begin{equation}
\Delta _{\mathrm{VK}}\equiv C_{0}\left( e^{3\lambda }-3e^{\lambda }\right)
\pm 2\sqrt{C_{d}^{2}-\kappa ^{2}}>0,  \label{Delta}
\end{equation}%
cf. Eq. (\ref{lambda0}). Straightforward analysis demonstrates that the
family of the fundamental solitons in the loosely knit lattice, with $%
C_{0}<C_{0}^{(0)}$, satisfies the VK criterion at all values of $\lambda $
where such solitons exist, i.e., in the region defined by Eq. (\ref%
{lambda_min}). On the other hand, for the family of dipole solitons, as well
as for the fundamental modes in the tightly knit lattice [see Eq. (\ref{C0}%
)], the $P(\lambda )$ dependence always has a minimum at $\lambda =\lambda
_{0}$, corresponding to $\Delta _{\mathrm{VK}}(\lambda _{0})=0$, see Eqs. (%
\ref{Delta}) and (\ref{lambda0}). The slope of the dependence is positive,
suggesting the VK stability, at $\lambda >\lambda _{0}$, and negative,
i.e.,VK-unstable, at $\lambda <\lambda _{0}$.

\begin{figure}[tbp]
\centering{\label{fig44a}
\includegraphics[scale=0.19]{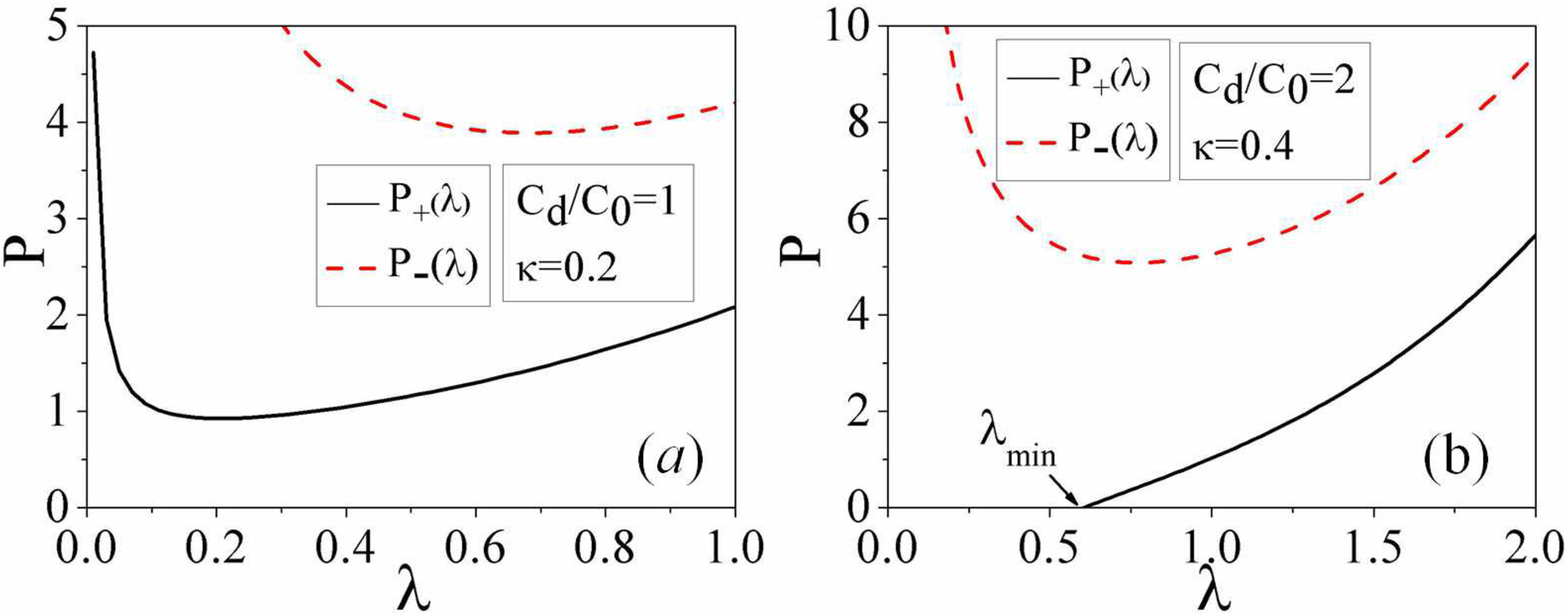}}%
\caption{Plots of the total power, $P_{+}(\protect\lambda )$
and $P_{-}(\protect\lambda )$ (black and dashed red curves, respectively),
defined as per Eq. (\protect\ref{P+-}), for tightly [$(\protect\kappa %
,C_{d})=(0.2,0.5)$] (a) and loosely [$(\protect\kappa ,C_{d})=(2,0.4)$] (b)
knit lattices, which are identified pursuant to Eq. (\protect\ref{C0}).
Recall that $P_{+}(\protect\lambda )$ and $P_{-}(\protect\lambda )$ pertain,
respectively, to the fundamental and dipole-mode discrete solitons.}
\label{PDelta}
\end{figure}

The above conclusions are illustrated by Fig. \ref{PDelta}, which displays $%
P_{\pm }(\lambda )$ for the fundamental solitons and dipole modes in the
tightly and loosely knit lattice, respectively. Below, we verify the
predictions of the VK criterion for the fundamental ($\phi _{+}$) and dipole
($\phi _{-}$) discrete $\mathcal{PT}$ solitons by means of computing stability
eigenvalues, and using direct simulations of the perturbed evolution.

\subsection{Numerical results for fundamental solitons}

The numerical analysis was started with search for stationary discrete
solitons by means of the imaginary-time propagation method \cite%
{Chiofalo,Jianke2}, with the purpose to check that there are no other
solutions but those found above in the analytical form, see Eqs. (\ref%
{analyt}), (\ref{DR}), (\ref{phi}), and (\ref{A^2}). In this way, no
additional solutions have been found, while the actually produced ones are
fully identical to their analytical counterparts.

For the fundamental solitons, which correspond to $\phi _{+}$ and $A_{+}^{2}$
in Eqs. (\ref{phi}) and (\ref{A^2}), the VK-stable region, as predicted by
Eq. (\ref{Delta}), is displayed in the plane of $(\kappa ,\lambda )$, for
different values of $C_{d}$, in Fig. 5(a1)-(a4). In these figures, the
VK criterion does not hold [i.e., $\Delta _{\mathrm{VK}}$ is negative, see
Eq. (\ref{Delta})] in gray areas. As this criterion is only necessary, but
not sufficient, for the stability, we have used numerically generated
solutions of the eigenvalue problem, based on Eq. (\ref{eigen}), to
identified fully stable areas [red in Fig. 5(a1)-(a4)], and those
(yellow ones) where the fundamental solitons are unstable, although they
obey the VK criterion.

\begin{figure}[tbp]
\centering{\label{fig3a}
\includegraphics[scale=0.18]{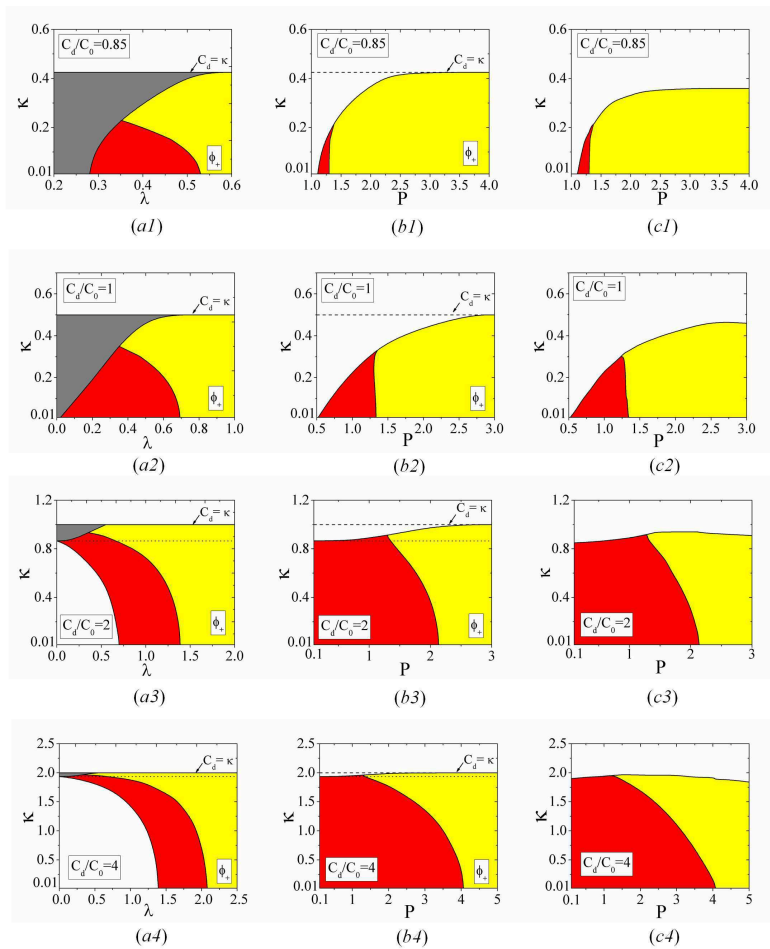}}%
\caption{(a1)-(a4) The existence and stability diagram for the
fundamental solitons in the plane of $\left( \protect\lambda ,\protect\kappa %
\right) $ for $C_{d}/C_{0}=0.85$ (a1), $C_{d}/C_{0}=1$ (a2), $C_{d}/C_{0}=2$
(a3), and $C_{d}/C_{0}=4$ (a4) [recall that $\protect\lambda $ determines the
localization of the solitons as per Eqs. (\protect\ref{analyt}) and (\protect
\ref{DR}), $\protect\kappa $ is the gain/loss coefficient of the $\mathcal{PT%
}$-symmetric defect, and $C_{d}$ is the coefficient of the coupling
between the two cores forming the defect]. The solitons do not exist
in the white regions, \textit{viz}., exactly as follows from the
analytical solution, at $\protect\kappa >C_{d}$ (at the top
of the panels) and at $\protect\lambda <\protect\lambda _{\mathrm{min}}(%
\protect\kappa )$ [see Eq. (\protect\ref{lambda_min})], i.e., in the left
bottom corners of panels (a3) and (a4). In the gray area, the solitons exist but do not satisfy the VK
criterion [see Eq. (\protect\ref{Delta})], i.e., they are definitely
unstable. The computation of the stability eigenvalues, based on Eq. (%
\protect\ref{eigen}), demonstrates that the solitons are completely stable
in the red area, but unstable against perturbation modes not covered by the VK criterion in the yellow
area. The white, gray and red/yellow colors have the same meaning in existence and stability diagrams
displayed in other figures below. The horizontal dotted lines in panel (a3)
and (a4) designate the border between regions of the\textit{\ tightly} and
\textit{loosely} knit linear lattices, defined as per Eq. (\protect\ref{C0}%
), the lattices being \textquotedblleft loose" [i.e., with $\protect\lambda %
_{\min }>0$, see Eq. (\protect\ref{lambda_min})] beneath these lines. In
panels (a1) and (a2), which correspond to $C_{0}\geq C_{d}$, the linear
lattices may only be tightly knit ones. (b1)-(b4) The stability and instability areas from (a1)-(a4) for the fundamental solitons, replotted in the
plane of $(P,\protect\kappa )$ at same values of $C_{d}$.(c1)-(c4) The same stability diagrams as in (b1)-(b4), but produced by direct simulations. In the white area, the imaginary-time simulations do not converge to stationary solitons.}
\label{Analregion}
\end{figure}

The stability regions from Fig. 5(a1)-(a4) are replotted in the $%
(P,\kappa )$ parameter plane in Fig. 5(b1)-(b4), as total power $P$
is a physically relevant characteristic of soliton modes. Further, Fig. 5(c1)-(c4) displays a counterpart of the same stability diagram, produced by
summarizing results of direct simulations of the perturbed evolution. The
comparison clearly shows that the computation of the eigenvalues, in the
combination with the VK criterion, and, on the other hand, direct
simulations lead to almost identical stability areas.

An evident trend featured by these results is destabilization of the
fundamental solitons with the increase of the total power. This can be
explained by the fact that, at large values of $P$, the nonlinearity, which
acts only at the two defect sites, becomes a dominant term, which makes the
system close to an isolated $\mathcal{PT}$-symmetric dimer \cite{dimer1,dimer2,dimer3,dimer4}, and
suppresses the stabilizing effect of the linear lattice attached to the
dimer.

In Fig. 5, the increase of the strength of the coupling between
the two defect-forming sites, $C_{d}$,$\ $makes the stability area much
larger. This feature can be readily understood, as, at small $C_{d}$, the
action of the gain at the amplified site is weakly checked by the power flow
to the adjacent dissipative site, hence perturbations can easily initiate
blowup in the gain-carrying core.

\begin{figure}[tbp]
\centering{\label{fig6a}
\includegraphics[scale=0.18]{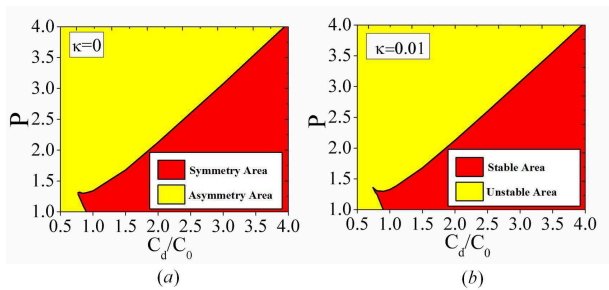}}%
\caption{(a) Areas of the stability (red) and instability
(yellow) for the fundamental discrete solitons in the $(P,C_{d}/C_{0})$
plane at a fixed value of the gain-loss coefficient, $\protect\kappa =0.01$.
(b) Regions of the existence of stable symmetric (red) and asymmetric
(yellow) fundamental discrete solitons in the same system with $\protect%
\kappa =0$ (no gain and loss).}
\label{PCdplane}
\end{figure}

The enhancement of the stability with the increase of $C_{d}$ is
additionally illustrated by Fig. 8(a), which shows the stability
areas of the fundamental solitons in the plane of $(P,C_{d}/C_{0})$ for a
small fixed value of the gain/loss coefficient of the $\mathcal{PT}$%
-symmetric defect, $\kappa =0.01$. From this figure, we conclude that the
stable fundamental solitons exist above the corresponding minimum value of
the coupling constant, $C_{d}/C_{0}>0.74$. At $P>1.5$, the stability
boundary may be fitted by a simple linear expression, $P=C_{d}/C_{0}$.
Further, in Fig. 8(b), we depict similar results for $\kappa =0$,
when the system reduces to the usual nondissipative one, with a double
nonlinear defect embedded into the linear array. Recall that both symmetric
\cite{embedded1,embedded2} and asymmetric \cite{asymm} discrete solitons can be
supported in the linear lattice by the double nonlinear defect. It is seen
in Fig. \ref{PCdplane} that the stability and instability areas for the
fundamental solitons in the $\mathcal{PT}$-symmetric system smoothly carry
over, respectively, into stability regions for symmetric and asymmetric
solitons in the nondissipative system. Indeed, because stationary asymmetric
solitons cannot exist in the $\mathcal{PT}$-symmetric system (as the balance
between the gain and loss cannot be maintained by them), the stability
region for asymmetric solitons in the conservative system, where symmetric
solitons exist too but are unstable \cite{asymm}, corresponds to the
instability region in the $\mathcal{PT}$-symmetric counterpart of the
conservative lattice.

\subsection{Numerical results for dipole solitons}

\begin{figure}[tbp]
\centering{\label{fig8a}
\includegraphics[scale=0.2]{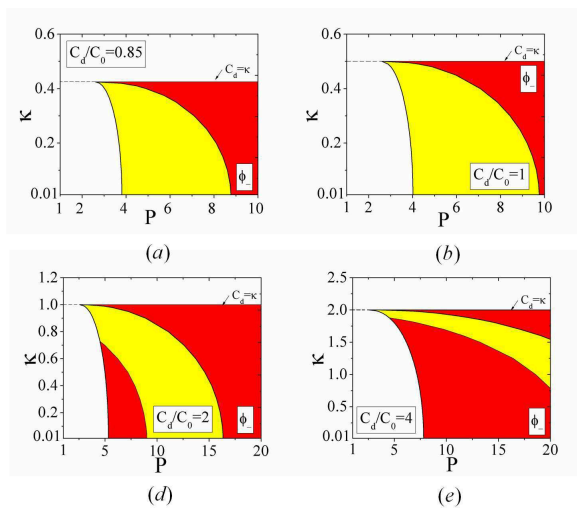}}%
\caption{Stability regions for the dipole-mode solitons with $%
C_{d}=0.85$ (a), $C_{d}=1$ (b), $C_{d}=2$ (c) and $C_{d}=4$ (d) in the $(P,%
\protect\kappa )$ plane. The meaning of the colors is the same as in other
figures: red for stable, and yellow for unstable.}
\label{PKanalytregionDipole}
\end{figure}

Recall that dipole solitons correspond to $\phi _{-}$ and $A_{-}^{2}$ in
Eqs. (\ref{phi}) and (\ref{A^2}), and they always have a nonzero existence
threshold in terms of the total power (unlike the fundamental solitons,
whose existence threshold vanishes in the loosely-knit lattices, as shown
above). Stability regions for the dipole modes in the $(P,\kappa )$ plane
are displayed in Fig. \ref{PKanalytregionDipole}. The stability and
instability in this figure are identified through the computation of
eigenvalues in the framework of Eq. (\ref{eigen}). Comparing this to similar
results for the fundamental solitons, displayed in Fig. 5(b1)-(b4)%
, we conclude that, as well as in the that case, the stability areas expands
with the increase of the coupling constant $C_{d}$ which links the two
defect-forming cores. On the other hand, the minimum (threshold) value of
the total power, which is necessary for the existence of the dipole
solitons, is higher than it was found above for their fundamental
counterparts (in the tightly-knit lattice). This difference is quite
natural, as fundamental solitons, being more compact than dipoles, need less
power to build themselves. Further, on the contrary to the fundamental
modes, the dipoles tend to be more stable at large values of the total
power, $P$ (although Fig. \ref{PKanalytregionDipole} exhibits alternation of
stability and instability areas). This trend may be understood via the
comparison of the present system with an isolated $\mathcal{PT}$-symmetric
dimer \cite{dimer1,dimer2,dimer3,dimer4}, to which it becomes closer with the enhancement of the
nonlinear terms acting solely at the defect sites.

\section{Scattering of incident waves on the defect}

\begin{figure}[tbp]
\centering{\label{fig10a} \includegraphics[scale=0.15]{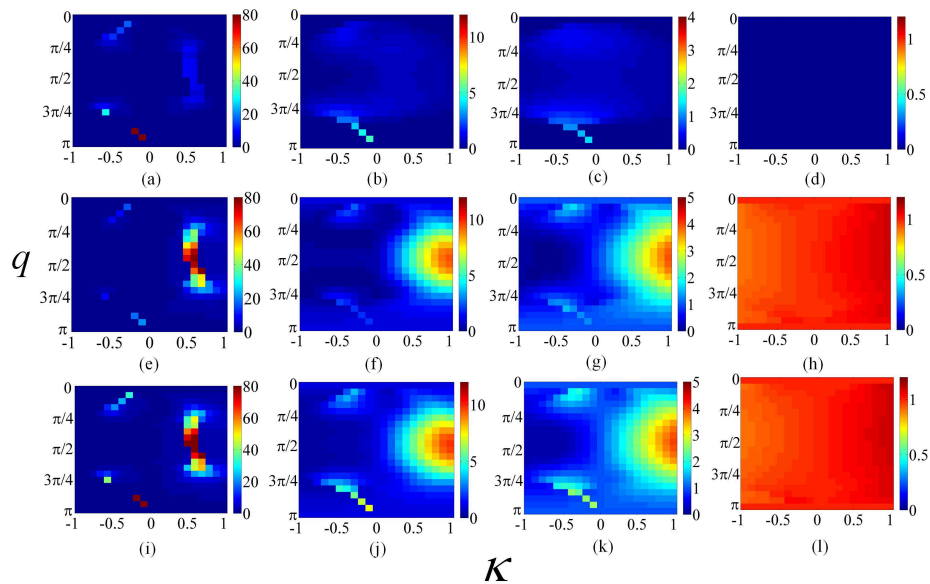}}
\caption{Numerically obtained solutions of the scattering
problem are shown in the plane of the gain-loss coefficient, $\protect\kappa
$, and wavenumber of the incident wave, $q$. Panels (a,b,c,d) show the
reflection coefficient, $|R|^{2}/I^{2}$, for $I=0.1$ (a), $0.5$ (b), $1$ (c)
and $10$ (d). (e,f,g,h): The transmission coefficient, $|T|^{2}/I^{2}$, for $%
I=0.1$ (e), $0.5$ (f), $1$ (g) and $10$ (h). (i,j,k,l): The sum of
reflection and transmission coefficients, $(|R|^{2}+|T|^{2})/I^{2}$, for $%
I=0.1$ (i), $0.5$ (j), $1$ (k) and $10$ (l). In these figures, we fix $%
C_{d}=C_{0}=0.5$. }
\label{ScatteringFig}
\end{figure}

In the framework of Eq. (\ref{DNLSE}), the scattering problem for linear
lattice waves with wavenumber $q$ hitting the embedded defect corresponds to
stationary solutions (\ref{uU}) with field $U_{n}$ looked for as
\begin{equation}
U_{n}=%
\begin{cases}
Ie^{iq(n-N/2)}+Re^{iq(n-N/2)}, & n\leq N/2 \\
Te^{iq[n-(N/2+1)]}, & n\geq N/2+1.%
\end{cases}
\label{stationaryscatter}
\end{equation}%
Here, $K$ and $q$ are related by the dispersion equation for propagating
waves,
\begin{equation}
K=2\cos (2q),  \label{dispersion}
\end{equation}%
cf. Eq. (\ref{DR}), $I$, $R$ and $T$ being amplitudes of the incident,
transmitted, and reflected waves, respectively. It is possible to fix $I$ as
a real positive constant, while $T=T_{r}+iT_{i}$ and $R=R_{r}+iR_{i}$ are
complex. Then, substituting the ansatz based on Eqs. (\ref{stationaryscatter}%
) and (\ref{dispersion}) into Eqs. (\ref{DNLSE}) at $n=N/2$ and $n=N/2+1$,
the following two complex equations for amplitudes $T$ and $R$ are obtained:
\begin{eqnarray}
&&2C_{0}\cos (2q)(I+R)=C_{0}(e^{-iq}I+e^{iq}R)+C_{d}T-|T+R|^{2}(T+R)-i\kappa
(I+R)  \label{TR} \\
&&2C_{0}\cos (2q)T=C_{0}Te^{iq}+C_{d}(I+R)+|T|^{2}T+i\kappa T  \notag
\end{eqnarray}%
A numerical solution of these equations produces the reflection and
transmission coefficients, i.e., $|R|^{2}/I^{2}$ and $|T|^{2}/I^{2}$, as
well as their sum, $(|R|^{2})+|T|^{2})/I^{2}$, as a function of wavenumber $%
q>0$ and $\mathcal{PT}$ coefficient $\kappa $, are displayed by the $%
(q,\kappa )$ plane in Fig. \ref{ScatteringFig} for different values of $I$.
Note that the plane comprises both $\kappa >0$ and $\kappa <$ $0$, which
correspond, respectively, to the incident wave originally hitting the
amplifying or dissipative element, according to Eq. (\ref{kappa}).
Naturally, there is no symmetry between outcomes of the scattering for $%
\kappa >0$ and $\kappa <0$ (a similar asymmetry explains unidirectional
transmission in $\mathcal{PT}$-symmetric systems \cite{uni1,uni2,uni3}).

From the figures, we conclude that, for $I=0.1$, which correspond to an
almost linear system, the solutions features a highly nonconservative
behavior, with the fields being strongly attenuated or, sometimes,
amplified, in different parts of the $(q,\kappa )$ plane. With the increase
of $I$, the nonlinear term becomes a dominant one, effectively suppressing
the nonconservative effects, which are generated by the linear gain and loss
terms. In particular, at $I=10$, which correspond to a strong nonlinearity,
the incident wave is almost entirely transmit across the defect.

\section{Conclusion}
The objective of this work is to propose a simple experimentally relevant
system, based on the linear waveguide array, with an embedded nonlinear
defect in the form of the $\mathcal{PT}$ dipole. The system makes it
possible to obtain the full family of exact solutions for fundamental and
dipole solitons pinned to the defect. The fundamental solitons in tightly
knit lattices, as well as the dipole modes in all cases, exist above a
finite threshold value of the total power, while for the fundamental
solitons in loosely knit lattices the threshold is absent. The stability of
the solitons is partly predicted by the VK criterion, and is obtained in the
full form via the numerical computation of eigenvalues for perturbation
modes. The fundamental and dipole solitons tend to be stable at lower and
higher values of the total power (norm), respectively. The stability areas
strongly expand with the increase of the strength of the coupling between
the two waveguides forming the defect, which can be readily explained. {The
scattering problem for lattice waves impinging upon the defect was solved
numerically. It shows a highly non-conservative behavior in the quasi-linear
limit, while the strong nonlinearity suppresses the nonconservative features
in the scattering states.}

It may be interesting to extend the analysis for two-dimensional arrays with
an inserted defect in the form of a $\mathcal{PT}$-symmetric nonlinear
quadrimer, as the corresponding generalization of the one-dimensional dimer.
In this case, however, analytical solutions for modes pinned to the defect
are not available, hence the entire analysis should be performed in a
numerical form.

\begin{acknowledgments}

This work is supported by the National Natural Science Foundation of China
(Grant Nos.11104083, 11204089).

\end{acknowledgments}

%

\bibliographystyle{plain}
\bibliography{apssamp}

\end{document}